\documentclass[fleqn,usenatbib]{mnras}

\usepackage[T1]{fontenc}
\usepackage{ae,aecompl}

\usepackage{graphicx}	
\usepackage{amsmath}	
\usepackage{amssymb}	
\usepackage{gensymb}
\usepackage{color}

\title[Asteroid--NS collision model for repeating FRBs]{Investigation of the asteroid--neutron star collision model for the repeating fast radio bursts}

\author[Smallwood et al.]{
Jeremy L. Smallwood,\thanks{E-mail: smallj2@unlv.nevada.edu}
Rebecca G. Martin
and Bing Zhang
\\
Department of Physics and Astronomy, University of Nevada, Las Vegas, 4505 South Maryland Parkway, Las Vegas, NV 89154, USA
}

\date{Accepted XXX. Received YYY; in original form ZZZ}

\pubyear{2018}

\begin{document}
\label{firstpage}
\pagerange{\pageref{firstpage}--\pageref{lastpage}}
\maketitle

\begin{abstract}
The origin of fast radio bursts (FRBs) is still a mystery.  One model proposed to interpret the only known repeating object, FRB 121102, is that the radio emission is generated from asteroids colliding with a highly magnetized neutron star (NS). With $N$--body simulations, we model a debris disc around a central star with an eccentric orbit intruding NS. As the NS approaches the first periastron passage, most of the comets are scattered away rather than being accreted by the NS. To match the observed FRB rate, the debris belt would have to be at least three orders of magnitude more dense than the Kuiper belt. We also consider the rate of collisions on to the central object but find that the density of the debris belt must be at least four orders of magnitude more dense than the Kuiper belt. 
These discrepancies in the density arise even if (1) one introduces a Kuiper-belt like comet belt rather than an asteroid belt and assume that comet impacts can also make FRBs; (2) the NS moves $\sim 2$ orders of magnitude slower than their normal proper-motion velocity due to supernova kicks; and (3) the NS orbit is coplanar to the debris belt, which provides the highest rate of collisions.


\end{abstract}




\section{Introduction}
Fast radio bursts (FRBs) are bright transients of radio emissions with
millisecond outburst durations. Despite the rapid observational progresses
\citep{Lorimer2007,Keane2011,Thornton2013,burke2014,Spitler2014,Ravi2015,Petroff2015,Masui2015,Keane2016,Spitler2016,champion2016,DeLaunay2016,Chatterjee2017},
thus far we still do not know the origin(s) of these mysterious
bursts. 
There are about two dozen FRBs
with a known source. Of these, there has been only one repeating
source, FRB 121102 \citep{Spitler2016,Scholz2016,Law2017}. 
Due to their high
dispersion measures ($\sim 500$ -- $\sim 3000\, \rm cm^{-3}\, pc$)
\citep{Thornton2013,Petroff2016},
FRBs most likely originate at cosmological distances. The repeating FRB 121102
was discovered to be associated with a steady radio emission source and localized to be
in a star-forming galaxy at red shift $z=0.193$
\citep{Chatterjee2017,Marcote2017,Tendulkar2017}, firmly establishing the cosmological nature of FRBs at least for this source. The bursts of FRB 121102 are sporadic \citep{Scholz2016,Law2017}. \cite{Spitler2016} reported $17$
bursts recorded from this source, which suggests a
repetitive rate of $\sim 3$ bursts per hour during the active phase \citep{Palaniswamy2016}.
Recently,  \cite{Michilli2018} reported almost 100\% linear polarization of the radio burst emission from FRB 121102 with roughly a constant polarization angle within each burst as well as a high and varying rotation measure. 

There have been many ideas proposed in the literature to explain the repeating bursts from FRB 121102. Widely discussed models include super-giant pulses from pulsars \citep{Connor2016,Cordes2016} or young magnetars \citep{Katz2016,Metzger2017,Margalit2018}. \cite{Zhang2017} interpreted the repeating bursts from FRB 121102 as due to repeated interactions between a neutron star (NS) and a nearby variable outflow. \cite{Michilli2018} suggested that the the steady radio emission of FRB 121102 could be associated with a low-luminosity accreting super-massive black hole. As a result, the source of variable outflow can be this black hole. \cite{Zhang2018} showed that this model can interpret the available data satisfactorily. 

This paper concerns another repeating FRB model that attributes the repeating bursts as due to multiple collisions of asteroids onto a NS \citep{Dai2016}.
\cite{Geng2015} initially described a mechanism where asteroids/comets may impact a NS to produce FRBs. As the impactor penetrates the NS surface, a hot plasma fireball forms. The ionized material located interior to the fireball expands along magnetic field lines and then coherent radiation from the top of the fireball may account for the observed FRBs. Since the acceleration and radiation mechanism of ultra-relativistic electrons remains unknown, a more detailed model of an asteroid-NS impactor was proposed by \cite{Dai2016}, where a highly magnetized NS travels through an asteroid belt around another star. They suggested that the repeating radio emission could be caused from the NS encountering a large number of asteroids. During each NS-asteroid impact, the asteroid has a large electric field component parallel to the stellar magnetic field that causes electrons to be scattered off the asteroidal surface and accelerated to ultra-relativistic energies instantaneously. Furthermore, \cite{Bagchi2017} argued that the model can interpret both repeating (when the NS intrudes a belt) and non-repeating (when the NS possesses the belt itself) FRBs. Asteroid impacts on NS were among early models for gamma ray bursts \citep{Harwit1973,Colgate1981,vanBuren1981,Mitrofanov1990,Shull1995} and soft gamma-ray repeaters \citep{Livio1987,Boer1989,Katz1994,Zhang2000}.

Debris discs are thought to be the remains of the planet formation process \citep{Wyatt2012,Currie2015,Booth2017}. They are observed to be common around unevolved stars  \citep{Moro-mart2010,Ballering2017,anglada2017}.  Debris discs around white dwarfs have not been directly observed, but their existence is implied by the pollution of their atmospheres by asteroidal material, perhaps from a debris disc that survived stellar evolution \citep{Gansicke2006,Kilic2006,vonHippel2007,Farihi2009,Jura2009,Farihi2010b,Melis2010,Brown2017,Bonsor2017,xu2018,Smallwood2018b}. However,  the existence of debris discs around NSs  is more uncertain \cite[e.g.,][]{Posselt2014}. The pulsar timing technique has a high level of precision which allows for the detection of small, asteroid mass objects around millisecond pulsars \citep{Thorsett1992,Bailes1993,Blandford1993,Wolszczan1994,Wolszczan1997}. No asteroids have been confirmed by observations and even the detections of planets around pulsars are rare \cite[][]{johnston1996,Bell1997,Manchester2005,Kerr2015,Martin2016pulsars}. Although, \cite{Shannon2013} suggested that an asteroid belt, having a mass of about $0.05\, \rm M_{\oplus}$, may be present around pulsar B1937+21.

Putting aside whether collisions between asteroids and NSs can emit coherent radio emission with high brightness temperatures to interpret FRBs, 
here we only consider whether a NS passing through a
debris disc around another star, either a main-sequence star or a white dwarf,
is able to produce a collision rate 
to match the observed rate in the repeating FRB 121102 during the active phase. In
Section~\ref{analytic} we examine analytically the expected rate of
asteroid collisions for reasonable debris disc parameters.  In
Section~\ref{numerical}, we use $N$--body simulations to model a
binary system with a debris disc of asteroids around another star to
determine the tidal disruption rate on to the companion NS. We
then consider the case that the central object is also a NS and
investigate the impact rate on to it. Finally we draw our conclusions
in Section~\ref{conc}.


\section{Analytical collision rate for a neutron star traveling through an asteroid belt}
\label{analytic}

We follow the approach of \cite{Dai2016} to calculate the collision
rate of asteroids with a NS passing through an asteroid belt. This analytical approximation is only relevant for the first periastron approach of the NS. As shown later in Section~\ref{numerical}, numerical simulations allow us to test the collision rate over several periastron approaches and to model a system that represents a captured NS sweeping through a belt. \cite{Dai2016} considered a
NS sweeping through the inner edge of an asteroid belt at $2\, \rm
au$. Each asteroid collision may give rise to an FRB. The impact rate
is estimated as 
\begin{equation}
\mathcal{R}_a = \sigma_a \nu_* n_a,
\label{rate1}
\end{equation}
where $n_a$ is the number density of the belt, $\sigma_a$ is the impact cross section described by \cite{Safronov1972} given by
\begin{equation}
\sigma_a = \frac{4\pi G M R_*}{v_*^2},
\label{sigma}
\end{equation}
$\nu_*$ is the proper velocity of the NS, $R_*$ is the radius of
the NS, and $M$ is the mass of the NS. There are two parameters
that the rate depends sensitively on: the number density of asteroids
in the belt and the velocity with which the NS moves. We consider
reasonable values for each below.

\subsection{Number density}
\label{numberdensity}

We estimate the number density of asteroids in the belt by assuming that the density is spatially uniform over the belt. For a belt of width and thickness $\eta R_{\rm a}$ with an inner radius $R_{\rm a}$, the number density is
\begin{equation}
n_a = \frac{N_a}{2\pi \eta^2 R_a^3},
\label{numden}
\end{equation}
Taking the parameters of \cite{Dai2016} of $N_{\rm a}=10^{10}$, $\eta=0.2$ and $R_{\rm a}=2\,\rm au$, the number density is $n_{\rm a}=4.97 \times 10^9 \, \rm au^{-3}$. With these parameters the collision rate may be sufficiently high to explain the repeating FRB (see also Section~\ref{sec:rate}). For comparison, we estimate the number density of the asteroid belt and the Kuiper belt in the solar system by assuming a cylindrical volume with height determined by the inclination distribution of the asteroids and comets.

\subsubsection{Comparison to the Solar System asteroid belt}
\label{abelt}

If the total energy released during an FRB is solely due to the gravitational potential energy of the asteroid and not due to the magnetic field energy of the NS, then the mass of an asteroid needed to produce an FRB can be estimated as done by \cite{Geng2015}. The asteroid mass required to enable a FRB as it collides with the NS is about $5.4 \times 10^{17}\, \rm g$ \citep{Geng2015}, which is  in the range of observed asteroid masses \cite[$10^{16}$--$10^{18}\, \rm g$, e.g.,][]{Colgate1981}.

 The present-day asteroid belt extends from about $2.0\, \rm au$ to $3.5\,\rm au$ \citep{Petit2001}. We can estimate the number of asteroids that are above a mass required to produce a FRB from the main belt size frequency distribution given by \cite{Bottke2005}.
In their table 1, the number of main belt asteroids with a radius greater than $\sim 4\, \rm km$ is approximately $N \sim 2.3\times 10^4$. To estimate the number density of the asteroid belt, we assume a volume produced by the inclination distribution of the asteroid belt being uniformly distributed between $-30$ and $30$ degrees \citep{Terai2011}. Thus, the number density of asteroids that are massive enough to produce a FRB is $n_{\rm asteroid} \approx 2.75 \times 10^2\, \rm au^{-3}$.


\cite{Dai2016} considered a typical iron-nickel asteroid to have a mass of $m = 2\times 10^{18}\, \rm g$, which is an order of magnitude larger than the minimum mass required to produce an FRB found by \cite{Geng2015}. With $10^{10}$ asteroids, their belt has a total mass of $16.7\, \rm M_{\oplus}$. The mass of the present-day asteroid belt is about $5\times 10^{-4}\, \rm M_{\oplus}$ \citep{Krasinsky2002}. While this is thought to be only $1\%$ of the mass of the original asteroid belt \citep{Petit2001}, the mass would need to be over five orders of magnitude higher to reach this level. Furthermore, the mass of a debris disc decreases over time due to secular and mean-motion resonances with giant planets
\citep{Froeschle1986,Yoshikawa1987,Morbidelli1995,Gladman1997,Morbidelli1998,Bottke2000,PetitMorbidelli2001,Ito2006,Bro2008,Minton2011,Chrenko2015,Granvik2017,smallwood2018a,Smallwood2018b}. These resonant perturbations cause eccentricity excitation which causes collisional grinding, which reduces the mass of the belt over time \citep{Wyatt2008}. A debris belt undergoes significant changes as the star evolves. If a belt is located within $\sim 100\, \rm au$ of the central star, as the star loses mass the belt undergoes adiabatic expansion in orbital separation \citep{veras2013}. Since debris discs lose mass over time due to collisional grinding, an asteroid belt around a NS may not be sufficiently massive to provide enough collisions.

\subsubsection{Comparison to the Kuiper belt}
\label{Kuiper_belt}
 A Kuiper belt analog, that is much more extended in size than an asteroid belt, may be a better source for FRB causing collisions with a neutron star.  The current observed mass of the Kuiper belt ranges from $0.01\, \rm M_{\oplus}$ \citep{Bernstein2004} to $0.1 \, \rm M_{\oplus}$ \citep{Gladman2001}, but there is a mass deficit to explain how the Kuiper belt objects accreted at their present heliocentric locations. Thus, the mass estimated in the initial Kuiper belt may be as much as $\sim 10\, \rm M_{\oplus}$ \citep{Stern1996,Stern1997a,Stern1997b,Kenyon1998,Kenyon1999a,Kenyon1999b,Kenyon2004,Delsanti2006}. 
The current Kuiper belt extends from about $30\, \rm au$ to $50\,\rm au$ \citep{Jewitt1995,Wiessman1995,Dotto2003}.  The number of discovered comets is only a small fraction of the theoretical total.  The number of Kuiper belt objects that have a radius greater than $R_{\rm min}$  is 
\begin{equation}
N_{> R_{\rm min}} = \frac{K}{2}\bigg[ \bigg(\frac{R_0}{R_{\rm min}}\bigg)^{2} -1 \bigg] + \frac{2K}{7},
\label{eq::total_num}
\end{equation}
\citep{Holman1995, Tremaine1990},
where $R_0$ is the largest comet radius and $K$ is related to the total belt mass $M$ with
\begin{equation}
M = \frac{4\pi}{3}\rho R_0^3 K C,
\end{equation}
where $\rho$ is the comet density and the constant $C = 3$ \citep{Holman1995}. We assume an upper limit for the current mass of the Kuiper belt, $M=0.1 \, \rm M_{\oplus}$ \citep{Gladman2001}. We take $R_{\rm min}$ to be the minimum radius needed to produce a FRB. We assume a spherical cometary nucleus with density $\rho = 1 \rm \, g\, cm^{-3}$. With the critical mass required to produce a FRB being $5.4 \times 10^{17}\, \rm g$ \citep{Geng2015}, the minimum radius of the object is set at $R_{\rm min} \approx 5 \, \rm km$. Thus, from equation~(\ref{eq::total_num}) the total number of objects with a size large enough to produce a FRB is $N_{> R_{\rm min}}  = 8.38 \times 10^8$. If the inclination is uniformly distributed between $-10$ and $10$ degrees \citep{Gulbis2010}, then the number density of objects in the Kuiper belt that are large enough to create an FRB is roughly $n_{\rm Kuiper} \approx 1.2 \times 10^4\,\rm au^{-3}$.

Next we compare the estimated number density of the present-day Kuiper belt to the  estimated number density of the primordial Kuiper belt. In the Nice model the outer Solar system began in a compact state \cite[$\sim 5.5\, \rm au$ to $\sim 14\, \rm au$, e.g.,][]{Levison2008}, and eventually Jupiter and Saturn migrated inward to their present-day locations and Uranus and Neptune migrated outward. When Jupiter and Saturn crossed their mutual 2:1 mean-motion resonance, their eccentricities increased. This sudden jump in their eccentricities caused the outward migration of Uranus, Neptune, and the destabilization of the compact primordial Kuiper belt. The timescale for Jupiter and Saturn to cross the 2:1 resonance was from about $60 \, \rm Myr$ to $1.1\, \rm Gyr$ \citep{Gomes2005}. Thus, we can assume that the compact primordial Kuiper belt was stable during this period of time, which may give enough time for a NS to be captured and plummet through the compact disc. Based on the Nice model, the primordial Kuiper belt was compact ($15-30\, \rm au$) and had an initial mass of $\sim 10\, \rm M_{\oplus}$
\citep{Gomes2005,Levison2008,Morbidelli2010,Pike2017}.

Assuming that the mass for the primordial Kuiper belt is $M\sim 10\, \rm M_{\oplus}$, we find that the total number of objects that are capable of producing a FRB is $N_{> R_{\rm min}}  = 8.38 \times 10^{10}$. This calculation assumes that the comet distribution is equivalent to that of the current Kuiper belt. We estimate the number density of the primordial compact Kuiper belt to be $n_{\rm Kuiper,p} \approx 4.8\times 10^{6}\, \rm au^{-3}$, which is  about two orders of magnitude higher than the current Kuiper belt.

\subsubsection{Extrasolar debris discs}
\label{extrasolar_discs}
Next, we compare extrasolar debris disc architectures with the Solar system and the theoretical belt used by \cite{Dai2016}. There have been hundreds of extrasolar debris discs that have been discovered over the past couple decades \cite[e.g.][]{Wyatt2008}. Since the emission from debris discs are optically thin, observations using submillimeter continuum can be used to estimate the disc masses, with the caveat that large bodies are missed.  Since one cannot detect asteroid-sized objects in debris belts, the presence of dust is used as an indicator of total disc mass. The majority of dust in debris belts are produced from asteroid and comet collisions due to eccentricity excitations from orbital resonances. Thus, the dust mass can be used an a predictor of the total mass of the disc by 
\begin{equation}
\frac{M_{\rm pb}}{t_{\rm age}}\approx \frac{M_d}{t_{\rm col}},
\label{discmass}
\end{equation}
\cite[e.g.,][]{Chiang2009}, where $M_{\rm pb}$ is the mass of the largest parent body at the top of the collisional cascade, $t_{\rm age}$ is the age of the system, $M_d$ is the dust mass and $t_{\rm col}$ is the collisional lifetime. The mass of largest parent body can be used as the minimum mass of the disc because larger bodies may exist collisionless over $t_{\rm age}$ \cite[e.g.,][]{Dohnanyi1969}. 

 The dust mass residing within debris discs have been observed in a plethora of planetary systems. Depending on the size of the grains, dust masses have been observed to be in the range $10^{-6}\, \rm M_{\oplus}$ to $10^{-1}\, \rm M_{\oplus}$ \cite[e.g.,][]{Matthews2007,Su2009,Patience2011,Hughes2011,Matthews2014,jilkov2015,Kalas2015,Nesvold2017}. Exozodical dust is the constituent for hot debris discs and these dust environments have been detected around two dozen main-sequence stars \citep{Absil2009,2013Absil,Ertel2014}. \cite{Kirchschlager2017} analyzed nine out of the two dozen systems and found that the dust should be located within $\sim 0.01$--$1\, \rm au$ from the star depending on the luminosity and that the dust masses amount to only ($0.2$--$3.5)\, \times 10^{-9}\, \rm M_{\oplus}$.

To calculate the minimum mass of the discs discussed above, based on the observed disc dust mass (see equation~(\ref{discmass})), we would have to calculate the collisional lifetime which is outside the scope of this paper. The main point about discussing some of the observed disc dust masses is to compare that to the Kuiper belt, which has a dust mass of $(3$--$5) \times 10^{-7} \, \rm M_{\oplus}$ \citep{Vitense2012}. The reason why the dust mass is so low in the Kuiper belt is that the belt has reached a steady-state where the amount of dust being ejected equals the amount being injected. The observed debris discs may not be in a steady-state, thus some have up to 6 orders of magnitude more dust than the Solar system. From equation~\ref{discmass}, if the amount of dust is large and the collisional timescale is short, then this suggests that some extrasolar debris discs may be more massive than the Kuiper belt or the asteroid belt. \cite{Heng2011} estimated the total mass of the debris disc in the system HD 69830, based on  the dynamical survival models of \cite{Heng2010}, to be $3$--$4 \times 10^{-3}\, \rm M_{\oplus}$, several times more massive than our asteroid belt. \cite{Chiang2009} found that low mass limit of Fomalhaut's debris disc to be about $3\, \rm M_{\oplus}$, a order of magnitude more massive than the observed mass in the Kuiper belt.



\subsection{Neutron star velocity}
\label{velocity}
One well studied type of a NS is a radio pulsar. We use the measured pulsar velocities to represent the proper motion velocities of NSs. Identifying pulsar proper motions and velocities is critical in understanding the nature of pulsar and NS astrophysics. Applications of pulsar velocity measurements include determining the birth rate of pulsars \citep{Ankay2004}, further understanding supernova remnants \citep{Migliazzo2002} and the Galactic distribution of the progenitor population \citep{chennamangalam2014}, and for this work, calculating the collision rate of asteroids with a NS. Pulsar velocities are calculated by measuring their proper motions and distances. 

The origin of pulsars high velocities at birth, also known as their natal kick velocities, are thought to be driven by an asymmetrical explosion mechanism \cite[e.g.,][]{Lai2006,Wongwathanarat2013}. For a review on pulsar natal kick velocities, see \cite{Janka2017}. The observed supernova explosions are not spherically symmetric \citep{Blaauw1961,Bhattacharya1991,Wang2001}. Natal kick velocities have typical values of $200$--$500\, \rm  km\, s^{-1}$ and up to about $1000\, \rm km\, s^{-1}$, with a mean velocity of $400\, \rm km\, s^{-1}$ \cite[e.g.,][]{Cordes1993,Harrison1993,Lyne1994,Kaspi1996,Fryer1998,Lai2001,Arzoumanian2002,Chatterjee2005,Hobbs2005}. The large eccentricities that are observed in Be/X-ray binaries also suggest large kick velocities \citep{Brandt1995,Bildsten1997,Martin2009}. 


 The average observed pulsar velocity is several hundred $\rm km\, s^{-1}$ \cite[e.g.,][]{Bailes1990,Caraveo1999,Hobbs2005,Deller2012,Temim2017,Deller2018}. There have been several mechanisms put fourth to explain high natal velocities of pulsars. Asymmetric neutrino emission was thought to be a mechanism that could provide kick velocities up to $\sim 300\, \rm km\, s^{-1}$ \citep{Fryer2006} but this mechanism may be ruled out due to the dependence on a very large magnetic field ($> 10^{16}\,\rm G$) and nonstandard neutrino physics \cite[e.g,][]{Wongwathanarat2010,Nordhaus2010,Nordhaus2012,Katsudia2018}. Also, \cite{Harrison1975} suggested that the electromagnetic rocket effect from an off-centered dipole in a rapidly rotating pulsar can accelerate pulsars up to similarly high velocities. Another mechanism is non-radial flow instabilities, such as convective overturn and the standing accretion shock instability \citep{Foglizzo2002,Blondin2003,Foglizzo2006,Foglizzo2007,Scheck2008}, which are able to produce asymmetric mass ejections during supernova explosions which can produce natal velocities from $100\, \rm km \, s^{-1}$ to up to and even beyond $1000 \, \rm km \, s^{-1}$. Next we explore the collision rate of asteroids on a pulsar with a pulsar velocity of $100\,\rm km\, s^{-1}$ \citep{Blaes1993,Ofek2009,Li2016}. The low value leads to a larger cross section area for the collisions and hence the maximum value for the collision rate.

\subsection{Collision rate}
\label{sec:rate}
The collision rate given by equation~(\ref{rate1}) is estimated as
\begin{align}
\mathcal{R}_a 
& = 1.25\bigg(\frac{R_*}{10\, {\rm km}}\bigg)\bigg(\frac{M}{1.4\,M_{\odot}}\bigg)\bigg(\frac{\nu_{*}}{100\, {\rm km\, s^{-1}}}\bigg)^{-1}
\notag \\
& \quad \quad \times 
\bigg( \frac{n_a}{4.97\times10^{9}\, {\rm au^{-3}}}\bigg) \rm h^{-1}.
\label{coll_rate}
\end{align}
Instead of an asteroid belt, we use the primordial Kuiper belt to calculate this rate.  We set $n_a$ to equal the density of the primordial Kuiper belt, $n_{\rm Kuiper,p} = 4.8\times 10^{6}\, \rm au^{-3}$. 
 We estimate a collision rate of $0.0012\, \rm h^{-1}$, which is about three orders of magnitude less than the analytical rate calculated by \cite{Dai2016}, which requires an extremely high debris disc density and a low NS velocity.  Our analytical calculation suggests that this mechanism cannot produce a comet collision rate of $3\, \rm h^{-1}$, even in the extremely dense primordial Kuiper belt. In the next section, we explore if our analytical findings can be supported by numerical integrations.

Previous works used the tidal disruption radius to calculate collisions, instead, we use the impact radius associated with equation~(\ref{sigma}). 
\cite{Colgate1981} defined the break up radius due to tidal forces to be
\begin{align}
R_b & = \frac{\rho_0 r_0^2 G M}{s}^{-1/2} \notag \\
& = 2.22 \times 10^4 \bigg(\frac{m}{10^{18}\, {\rm g}}\bigg)^{2/9} \bigg( \frac{\rho_0}{8\times 10^{15}\, {\rm g\, km^{-3}}}\bigg) 
\notag \\
& \quad \quad \times \bigg(\frac{s_0}{10^{20}\, {\rm dyn\, km^{-2}}}\bigg)^{1/3} \bigg(\frac{M}{1.4\,M_{\odot}}\bigg)^{1/3}\, \rm km,
\end{align}
where $\rho_0$ is the density of the asteroid, $r_0$ is the cylindrical radius of the particle, and $s_0$ is the tensile strength. The impact radius is defined as
\begin{align}
R_{\rm Impact} & = \sqrt{\frac{4GMR_*}{v_*^2}} \notag \\
& = 2.73\times 10^4 \bigg(\frac{M}{1.4\,{\rm M_{\odot}}}\bigg)^{1/2} \bigg(\frac{R_*}{10\, \rm km}\bigg)^{1/2} 
\notag \\
& \quad \quad \times \bigg(\frac{v_*}{100\, \rm km}\bigg)^{-1} \, \rm km.
\end{align}
We find that the impact radius is larger than the tidal breakup radius. \cite{Dai2016} specifically required asteroids rather than comets to produce FRBs. This is because the size of the asteroid is small enough to produce a duration of order of milliseconds, which is consistent with the typical durations of FRBs. With a spherical comet nucleus with a radius $r_0 = 5\, \rm km$, the duration can be estimated as
\begin{equation}
\Delta t \simeq \frac{12r_0}{5}\bigg(\frac{2GM}{R_{\rm impact}}\bigg)^{-1/2},
\end{equation}
giving a duration of $3.3\, \rm ms$. This duration is consistent with the pulse width of FRB 121102, which is observed at $3 \pm 0.5 \, \rm ms$ \citep{Spitler2014}. However, this calculation just encompasses the cometary nucleus and neglects the cometary tail. A long cometary tail could potentially  destroy the coherent emission responsible for producing FRBs.



\section{$N$--Body Simulations}
\label{numerical}

\begin{figure*}
\includegraphics[width=8.7cm]{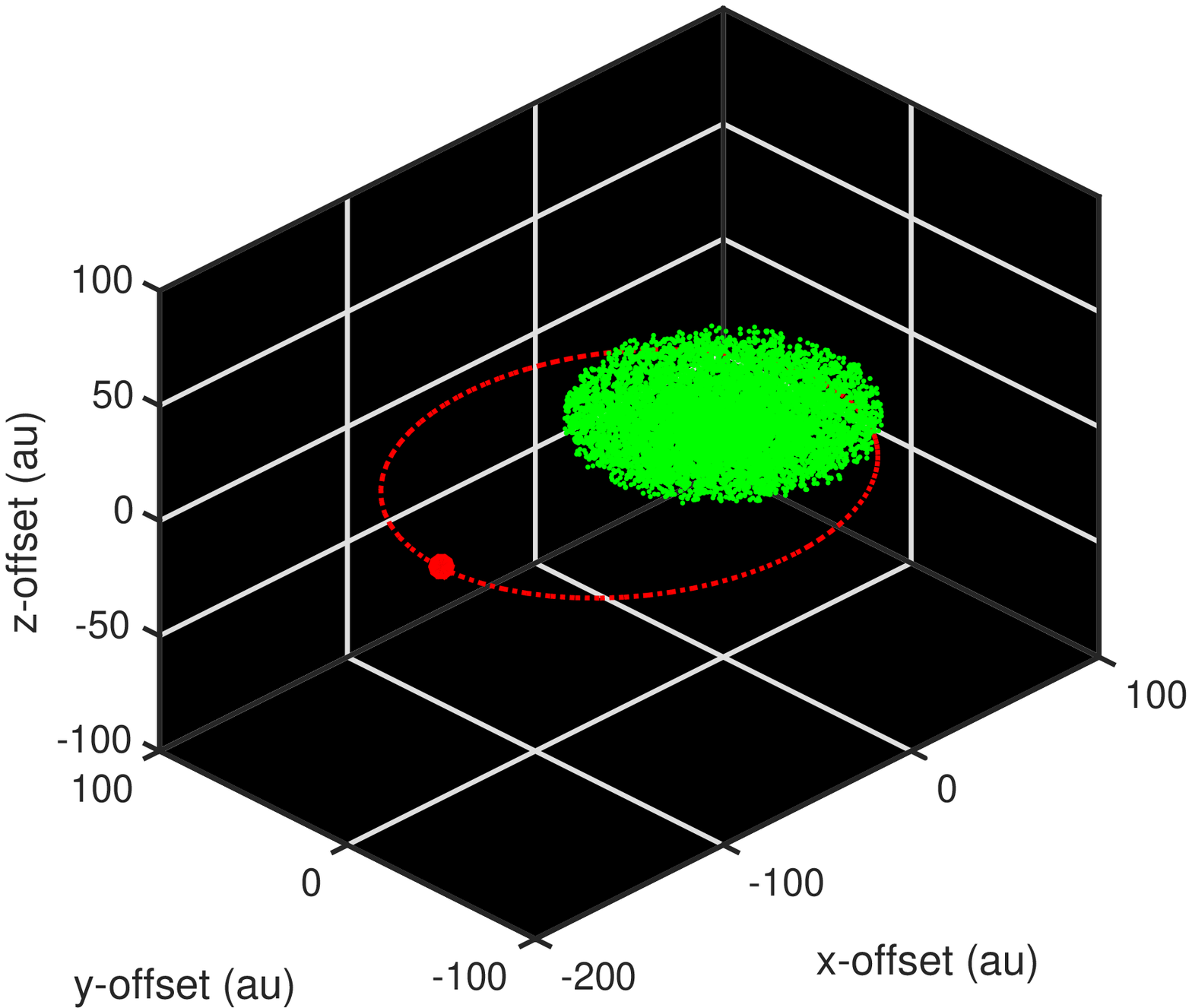}
\includegraphics[width=8.7cm]{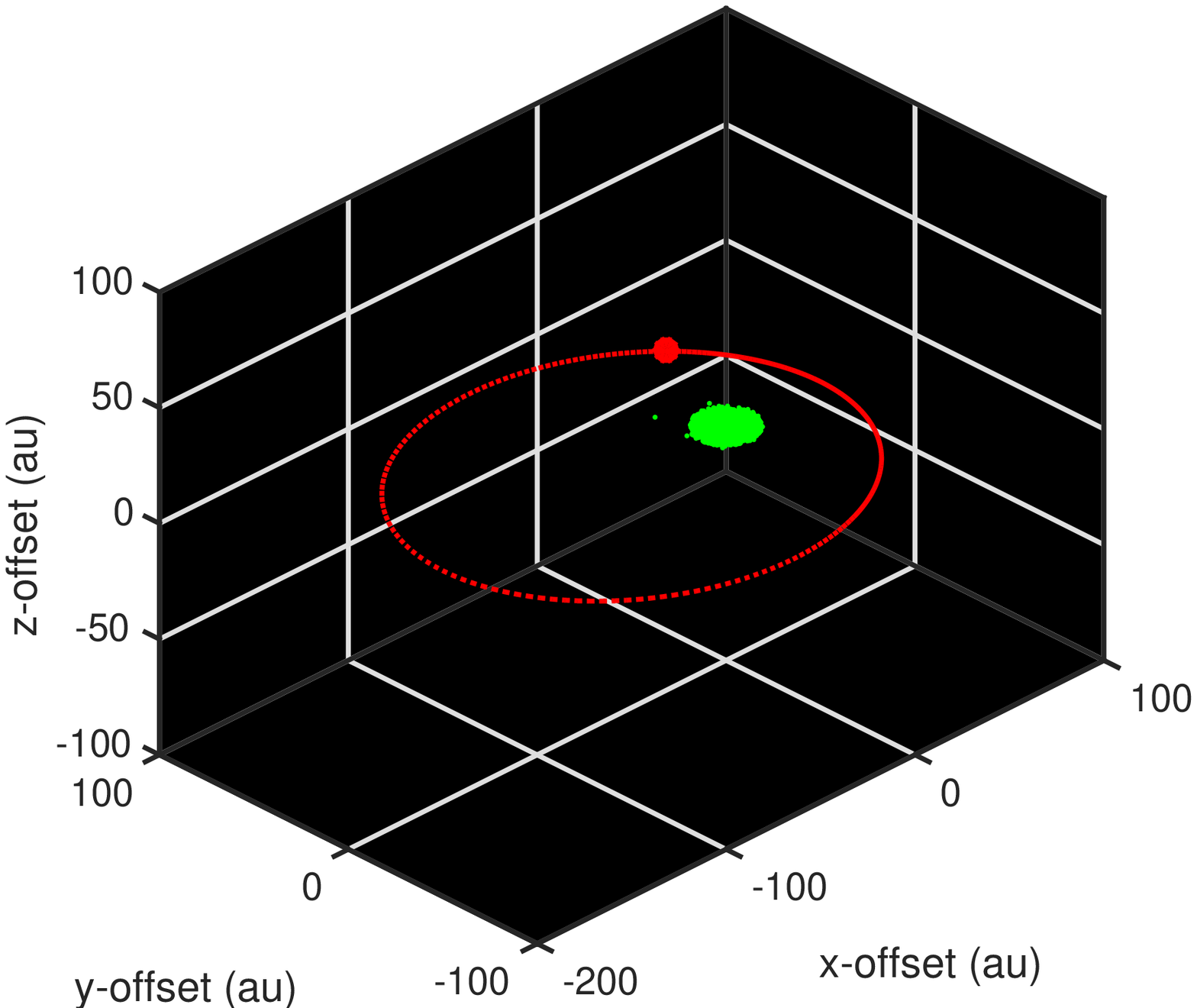}
\caption{Three-dimensional models of the debris disc distribution (shown by the green dots) at $t = 0\, \rm yr$ (initial distribution, left panel) and at $t = 100,000\, \rm yr$ (final distribution, right panel). The orbit of the NS that sweeps through the debris disc is shown by the dotted red curve, with the red dot signifying the position of the NS. The red dot has been inflated in order to visibly enhance the location. The NS has an eccentricity of $0.5$ and is initially located at apastron. The reference frame in centered on the central star (where the debris disc is orbiting), which is located at the origin (not shown).}
\label{setup}
\end{figure*} 

\begin{figure*}
\includegraphics[width=8.7cm]{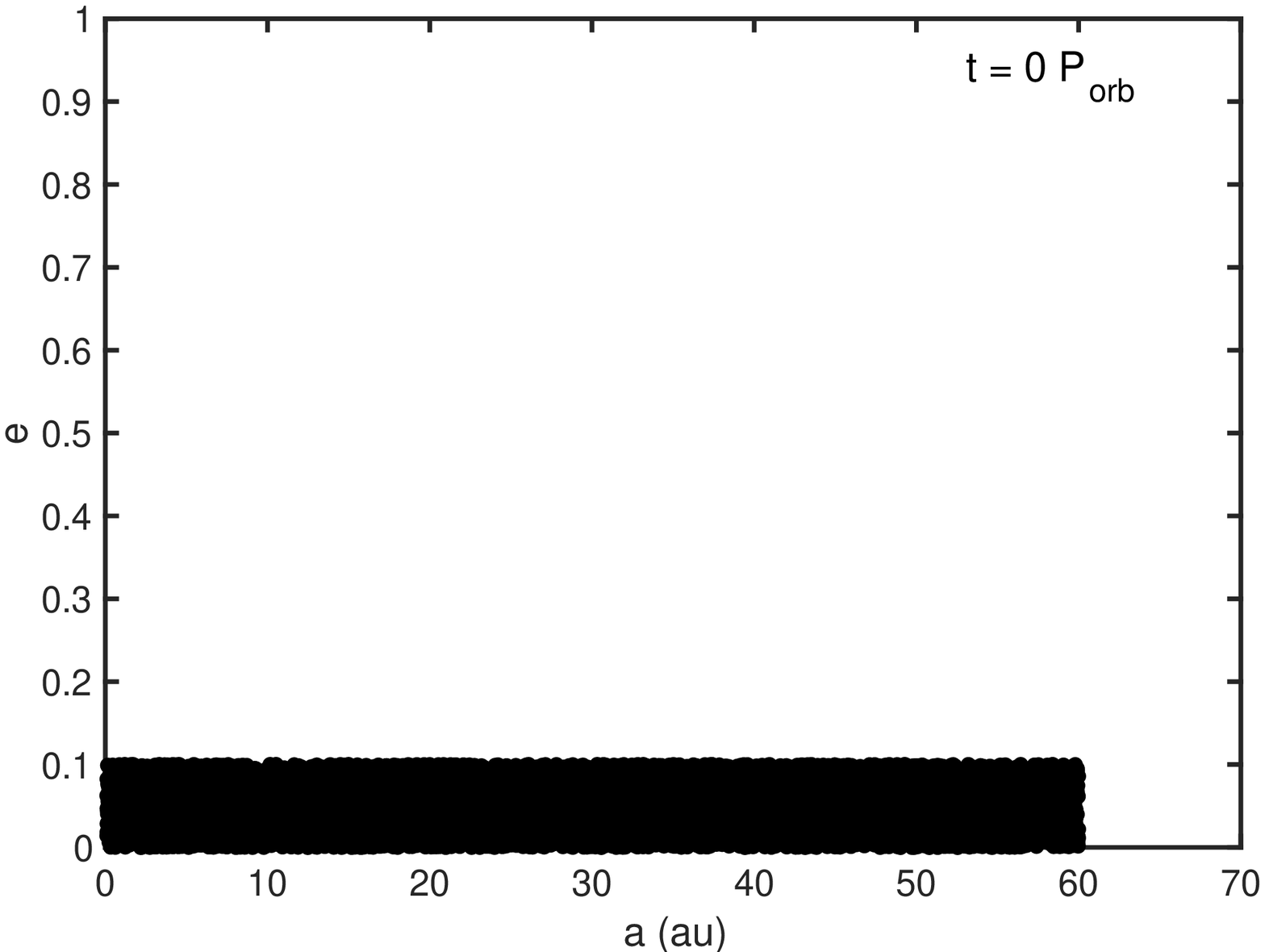}
\includegraphics[width=8.7cm]{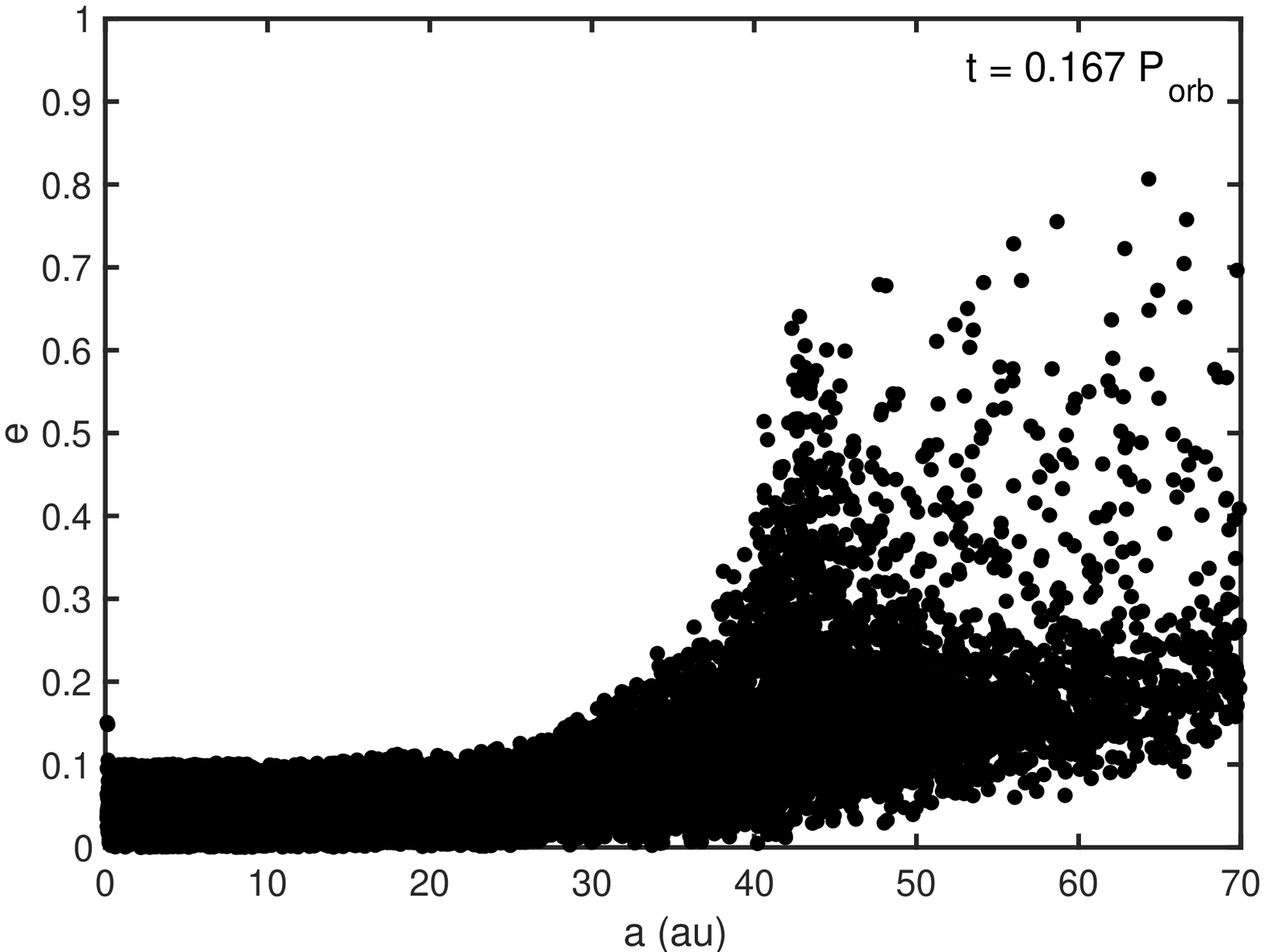}
\includegraphics[width=8.7cm]{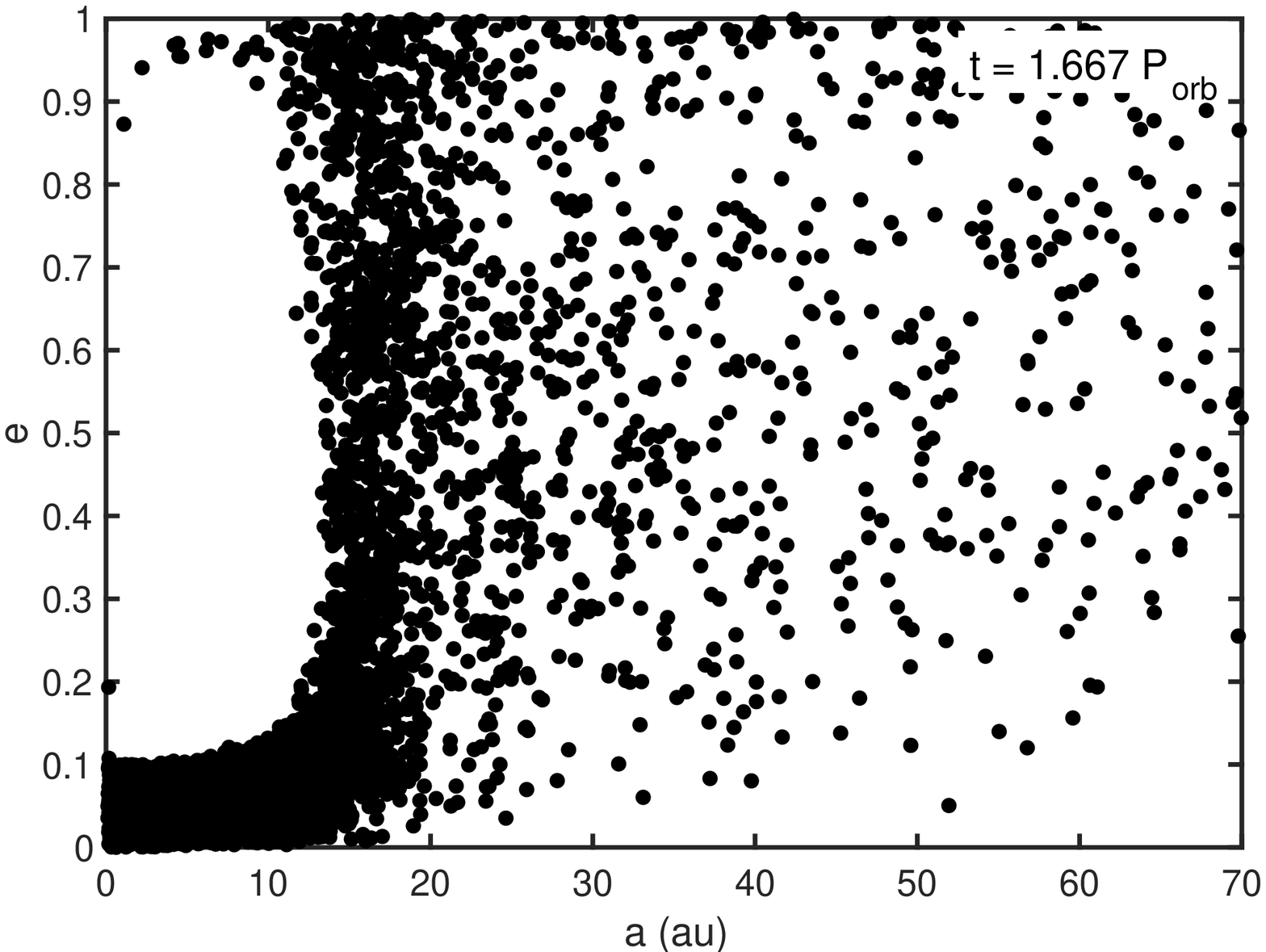}
\includegraphics[width=8.7cm]{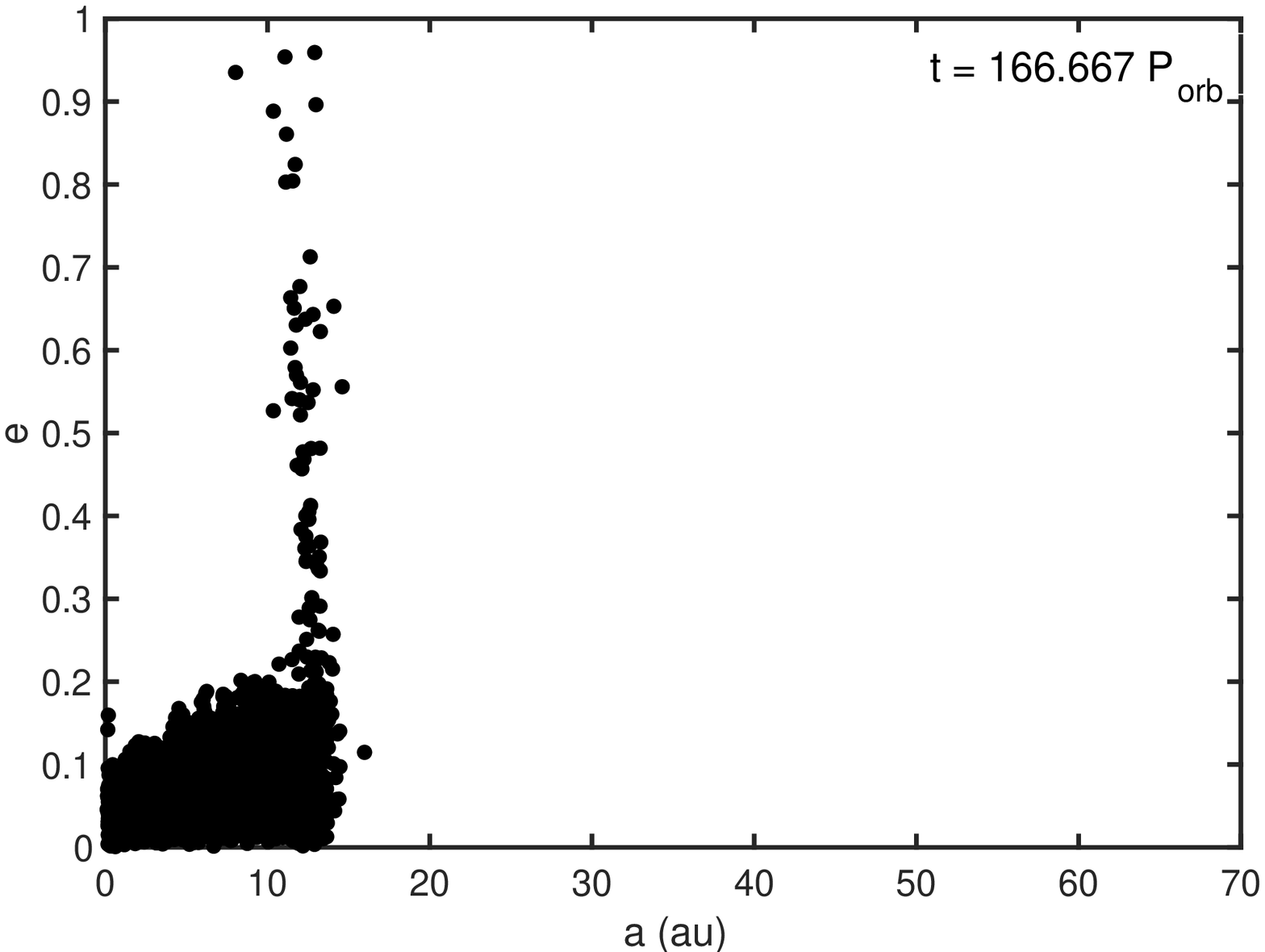}
\caption{The orbital eccentricity distribution of the fiducial debris disc an a function of semi-major axis at $t = 0\, P_{\rm orb}$ (top left panel), at $t = 0.67\, P_{\rm orb}$ (top right panel), at $t = 1.67\, P_{\rm orb}$ (bottom left panel), and at $t = 166.67\, P_{\rm orb}$ (bottom right panel). Initially, the orbiting NS begins at apastron and has an orbital period of about $600\, \rm yr$.}
\label{e0p5}
\end{figure*}

\begin{figure*}
\includegraphics[width=8.7cm]{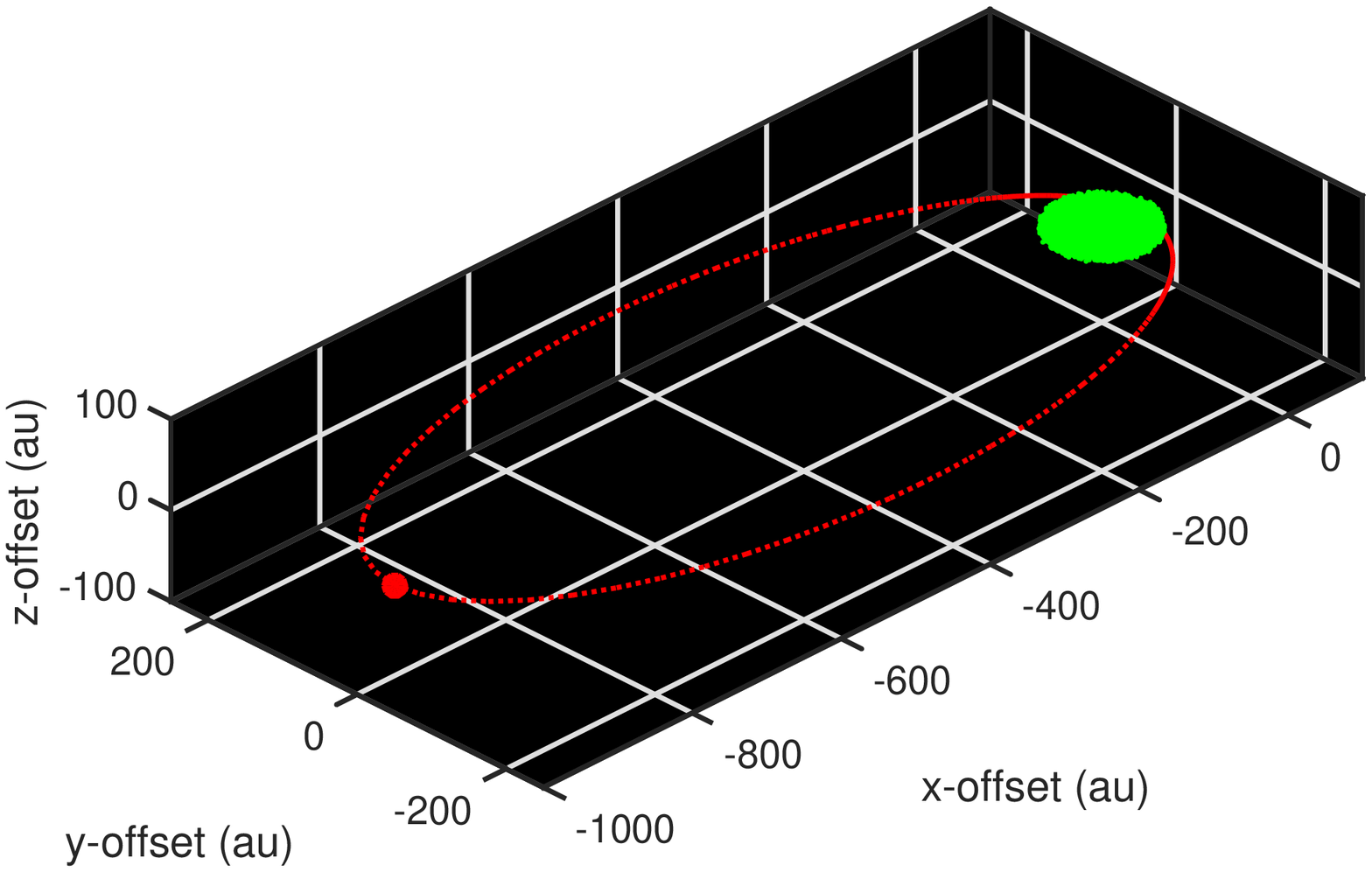}
\includegraphics[width=8.7cm]{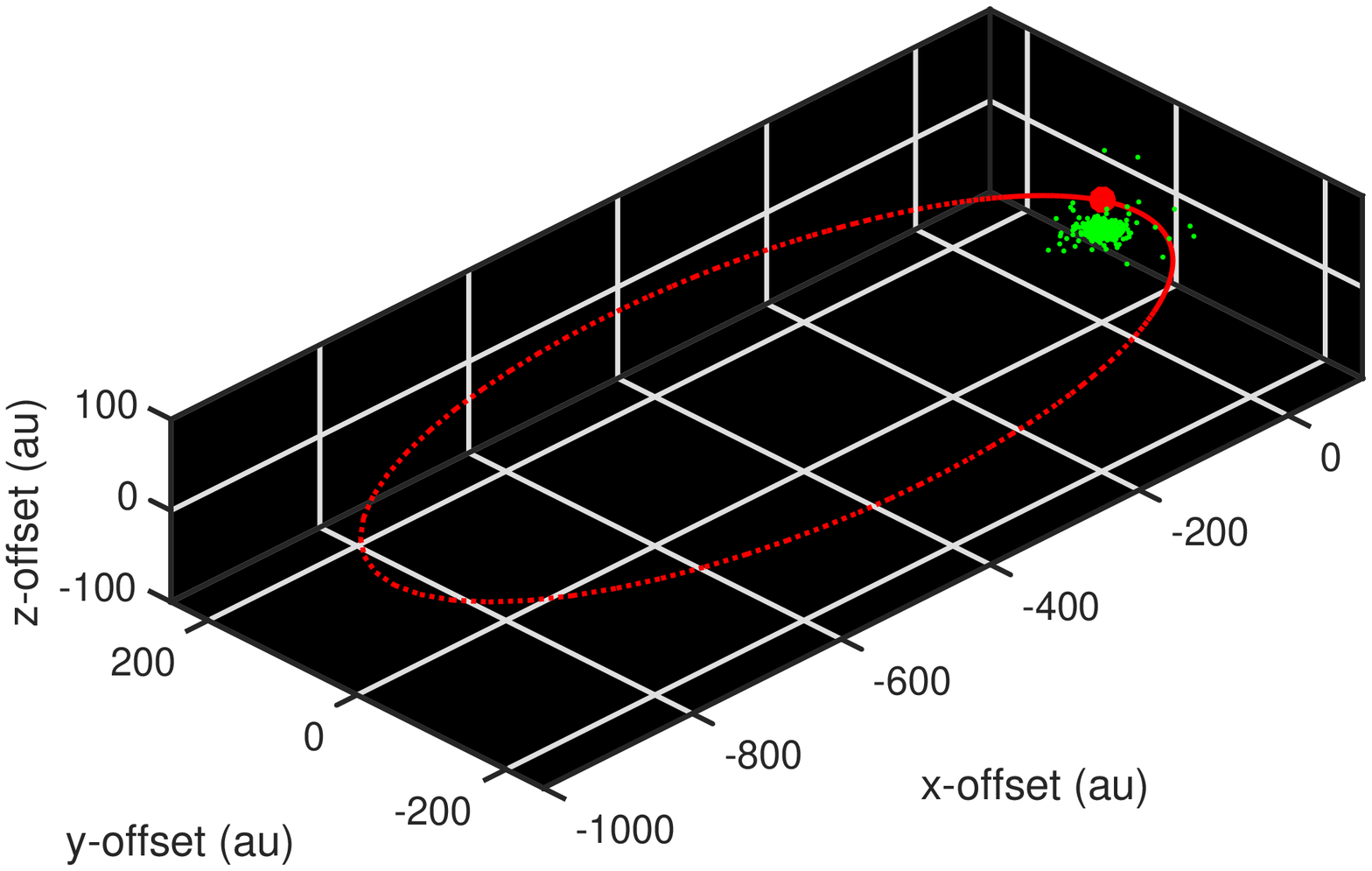}
\caption{Same as Fig.~\ref{setup} but with an eccentricity of $0.9$.}
\label{e0p9}
\end{figure*}

\begin{figure*}
\includegraphics[width=8.7cm]{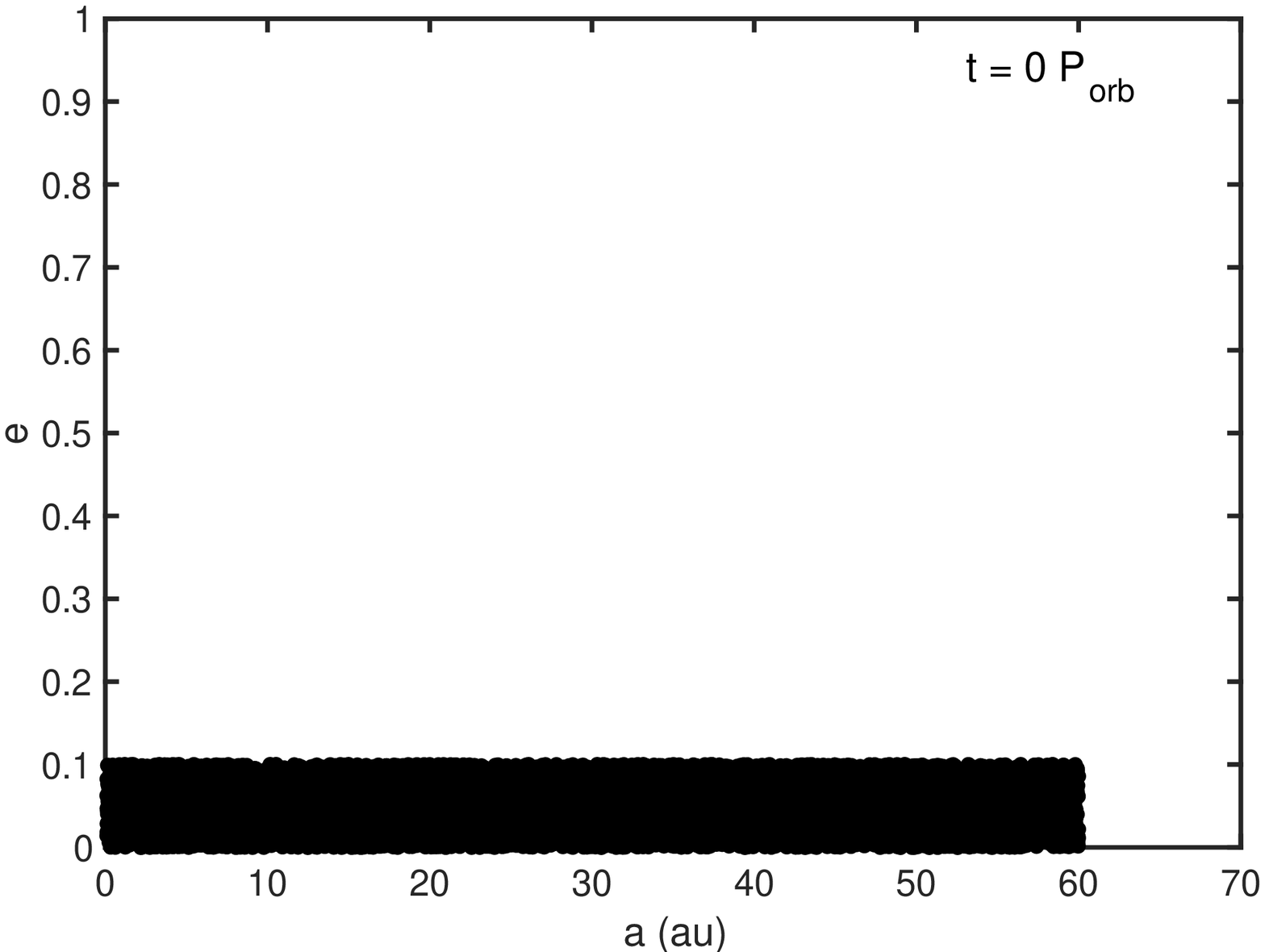}
\includegraphics[width=8.7cm]{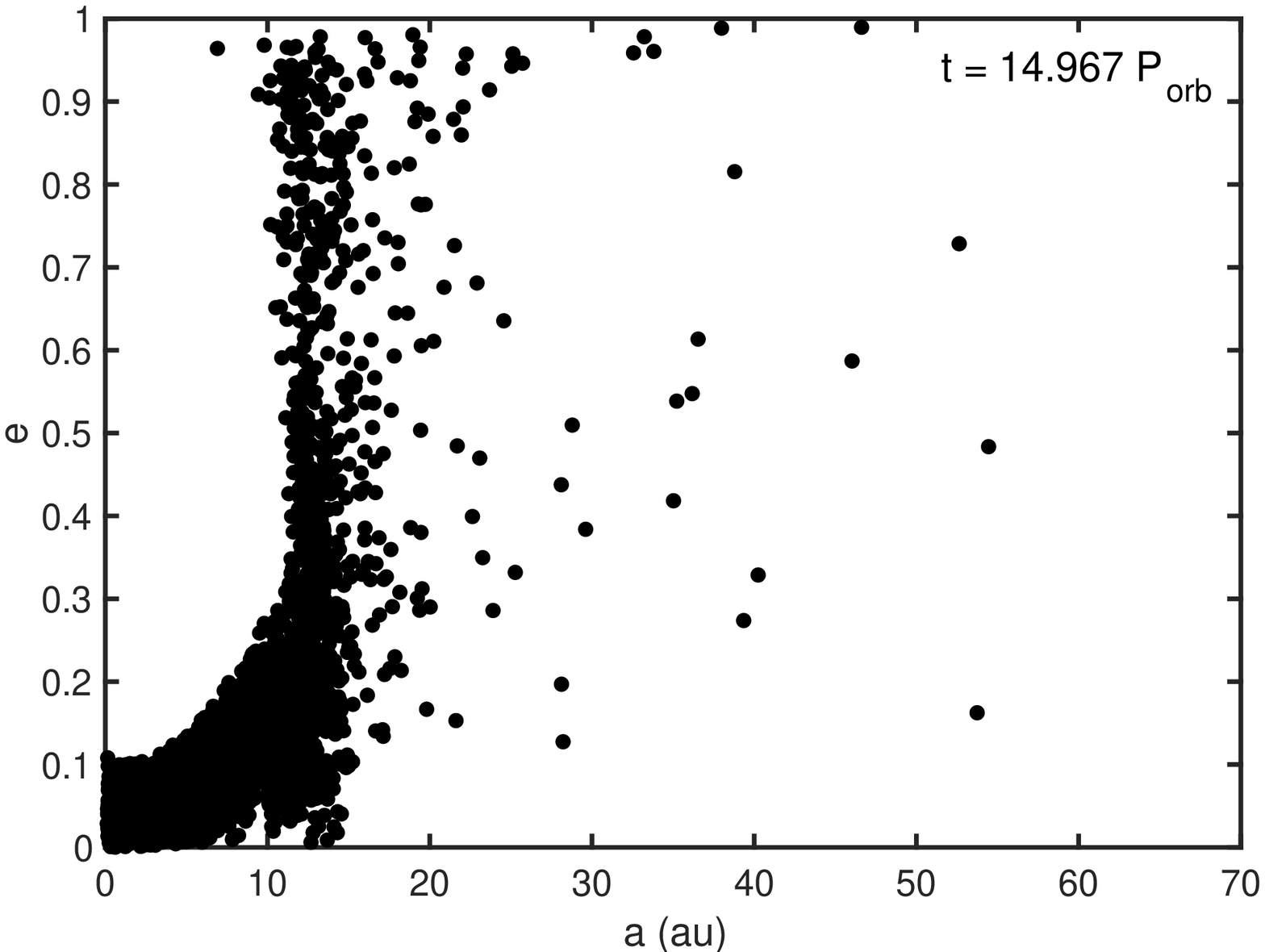}
\caption{Same as Fig.~\ref{e0p5} but with an eccentricity of $0.9$ and at times $t = 0\, P_{\rm orb}$ (left panel) and $t = 14.97\, P_{\rm orb}$ (right panel). Initially, the orbiting NS begins at apastron and has an orbital period of about $6681.5\, \rm yr$.}
\label{e0p9_dist}
\end{figure*}

We investigate whether asteroid/comet collisions can occur on a NS at a rate high enough to explain the repeating FRB 121102. We examine two scenarios, in the 
the first scenario, the NS formed in a binary. In the second scenario, the NS was captured into a binary. 

In the non-capture scenario, the NS orbit is coplanar to the debris disc, with an eccentricity of $e = 0.5$, a semimajor axis of $a = 100\, \rm au$,
and an orbital period of $P_{\rm orb} = 597.6\, \rm yr$. The
assumption of coplanarity gives the highest collision rate
possible. As the NS is formed from a supernova explosion, the
NS will receive a kick which can lead to an eccentric orbit
\citep{Blaauw1961,Bhattacharya1991}. In the capture scenario we also
assume coplanarity, along with an eccentricity of $e = 0.9$, a
semimajor axis of $a = 500\, \rm au$, and an orbital period of $P_{\rm
  orb} = 6681.5\, \rm yr$. Even though an eccentricity of $0.9$ is
technically bound, for simplicity, we assume that this eccentricity
resembles a capture.
In both scenarios we assume the binary system to
be of equal mass of $1.4\, M_{\odot}$, with the frame of reference
centered on the central star with the debris disc. We create a
Kuiper--belt like fiducial disc of $10,000$ test particles with the
orbital elements described as follows. The semimajor axis ($a$) is
randomly allocated in the range [0.1 60] au, the eccentricity ($e$) is
randomly distributed in the range [0 0.1], and the inclination ($i$)
is randomly selected in the range [0 10]\degree. The remaining
rotation orbital elements, the argument of pericenter ($\omega$), the
longitude of the ascending node ($\Omega$), and the mean anomaly
($\mathcal{M}$), are all randomly allocated in the range [0
  360]\degree. The NS companion begins at apastron.
  
Since in both cases, the intruding NSs are in bound orbits. We calculate the periastron velocities in both scenarios and compare that to the NS natal kick velocity used in the analytical approximation in equation~(\ref{coll_rate}). For the NS with eccentricities of $0.5$ and $0.9$, the periastron velocities are $6.1048\, \rm km/s$ and $6.8707\, \rm km/s$, respectively, with each having a periastron distance of $50\, \rm au$. These velocities are about two orders of magnitude lower than the average NS velocities, which means the number of collisions from the numerical results should be heightened due to the extremely low periastron velocity.

We model the NS system along with a debris disc using the $N$--body sympletic integrator in the orbital dynamics package, {\sc mercury} \citep{Chambers1999}. We simulate this system for a duration of $100,000$ years, which corresponds to a time of $166.67\, P_{\rm orb}$ for the non-capture scenario and a time of $14.97\, P_{\rm orb}$ for the capture scenario, where we calculate the number of test particles that impact the central star and the companion.  We physically inflate the radius of the NS and the central star to the impact radius. When a test particle collides with a either star it is considered to have been impacted and removed from the simulation.
The system is in a initial stable configuration without the intruding NS.

The left panel of Fig.~\ref{setup} shows the initial setup of the non-capture scenario. The orbit of the intruding NS that sweeps through the fiducial belt is shown by the red dashed line. The frame of reference is centered on the central star (which is not shown), which is located at the origin, $(0,0,0)$. The NS is initially at apastron, with the red dot being inflated in order to visibly enhance the location. {\sc Mercury} uses  the mean anomaly as one of the rotational elements. In order to construct the orbit of the NS in Fig.~\ref{setup}, we make use of the first-order transformation from mean anomaly to the true anomaly ($\nu$) given by
\begin{equation}
\mathcal{M} = \nu - 2e\sin \nu,
\end{equation}
where $e$ is the eccentricity of the NS.

\begin{figure*}
\includegraphics[width=8.7cm]{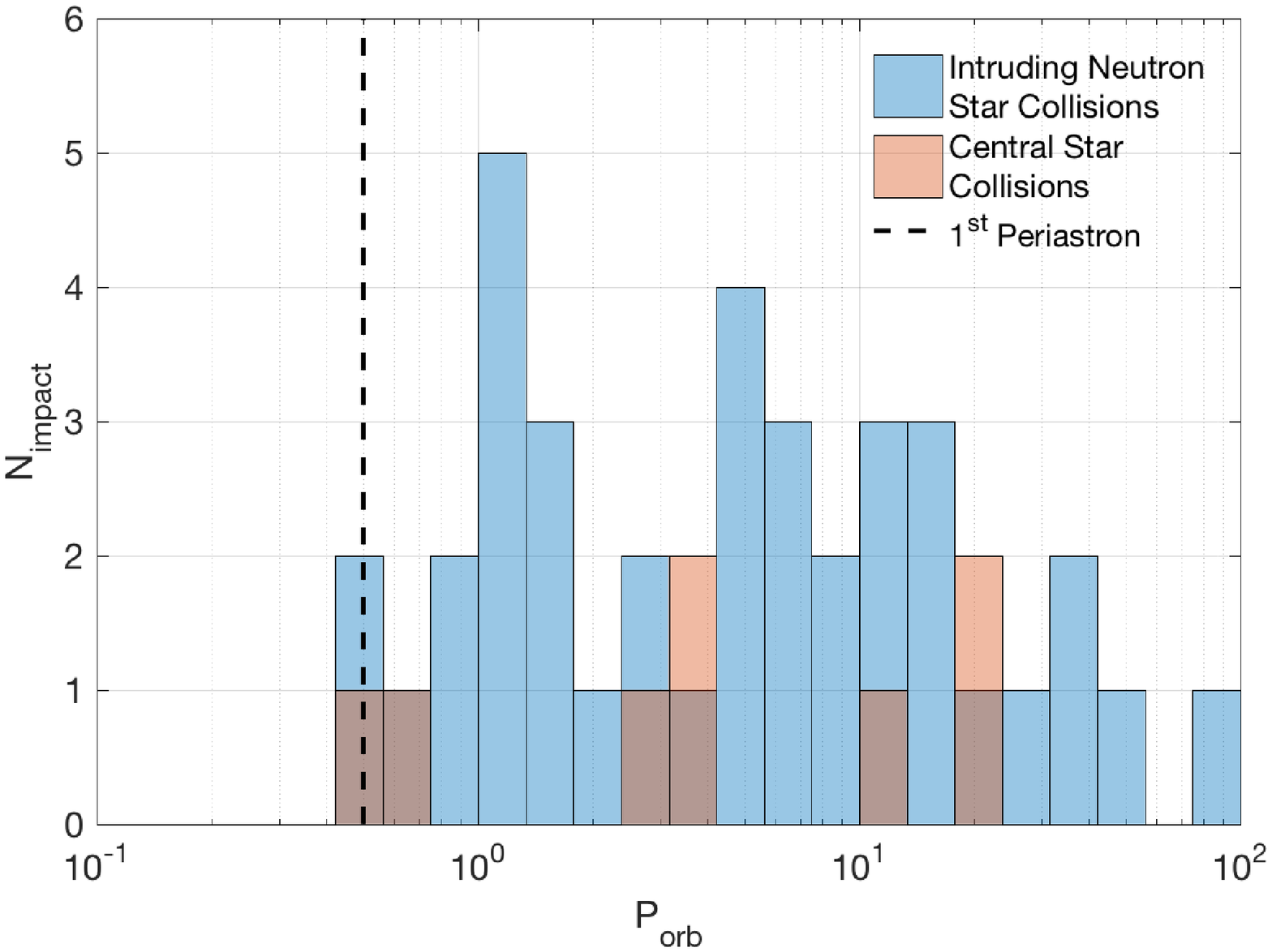}
\includegraphics[width=8.7cm]{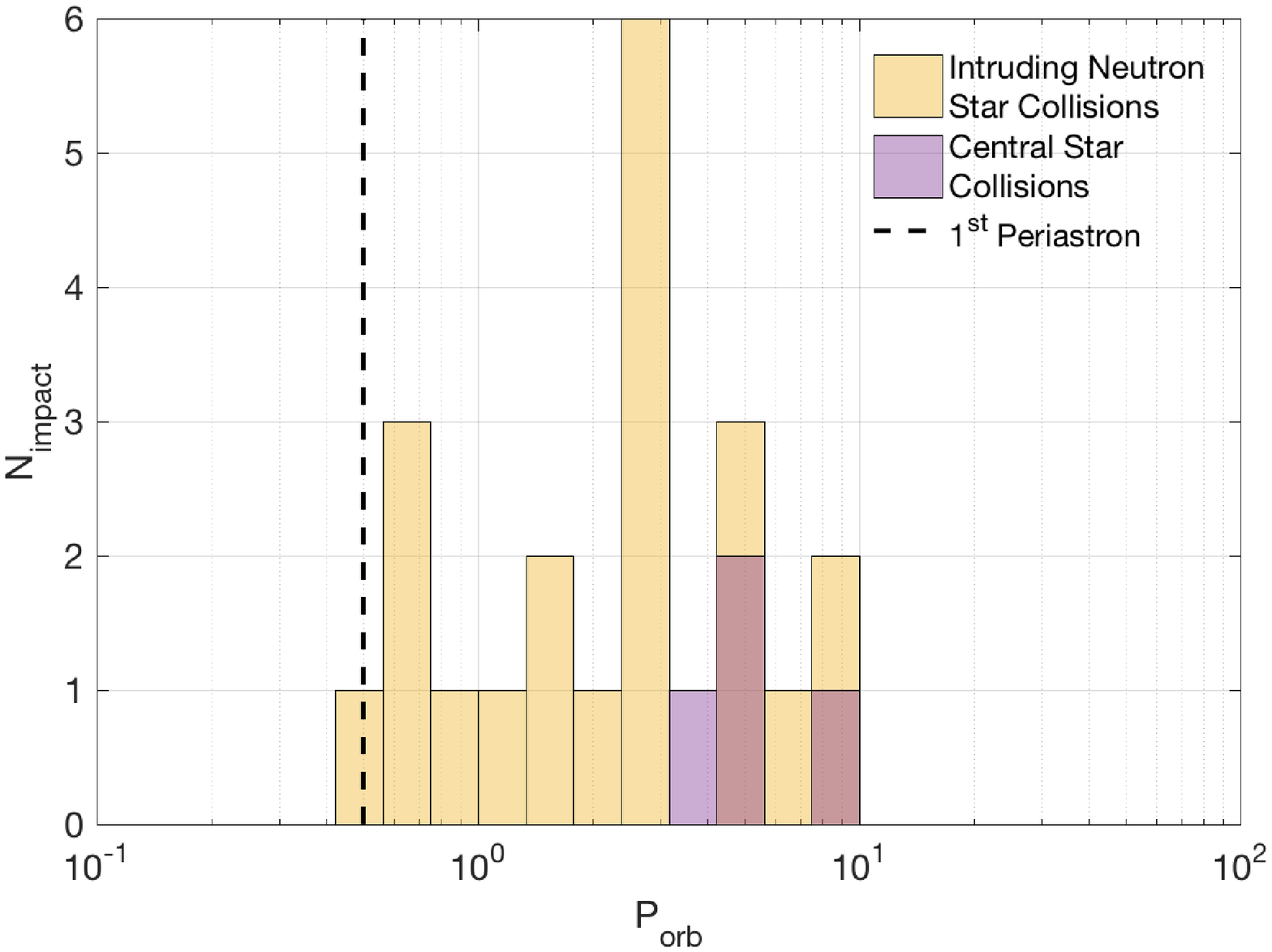}
\caption{The number of impact events as a function of time for the non-capture scenario (left panel) and for the capture scenario (right panel). In the left panel, the intruding NS is denoted with blue and central star with red. In the right panel, the intruding NS is denoted with yellow and central star with purple. The times of the first periastron passage are shown by the horizontal dotted lines.}
\label{rate}
\end{figure*}

\begin{figure*}
\includegraphics[width=8.7cm]{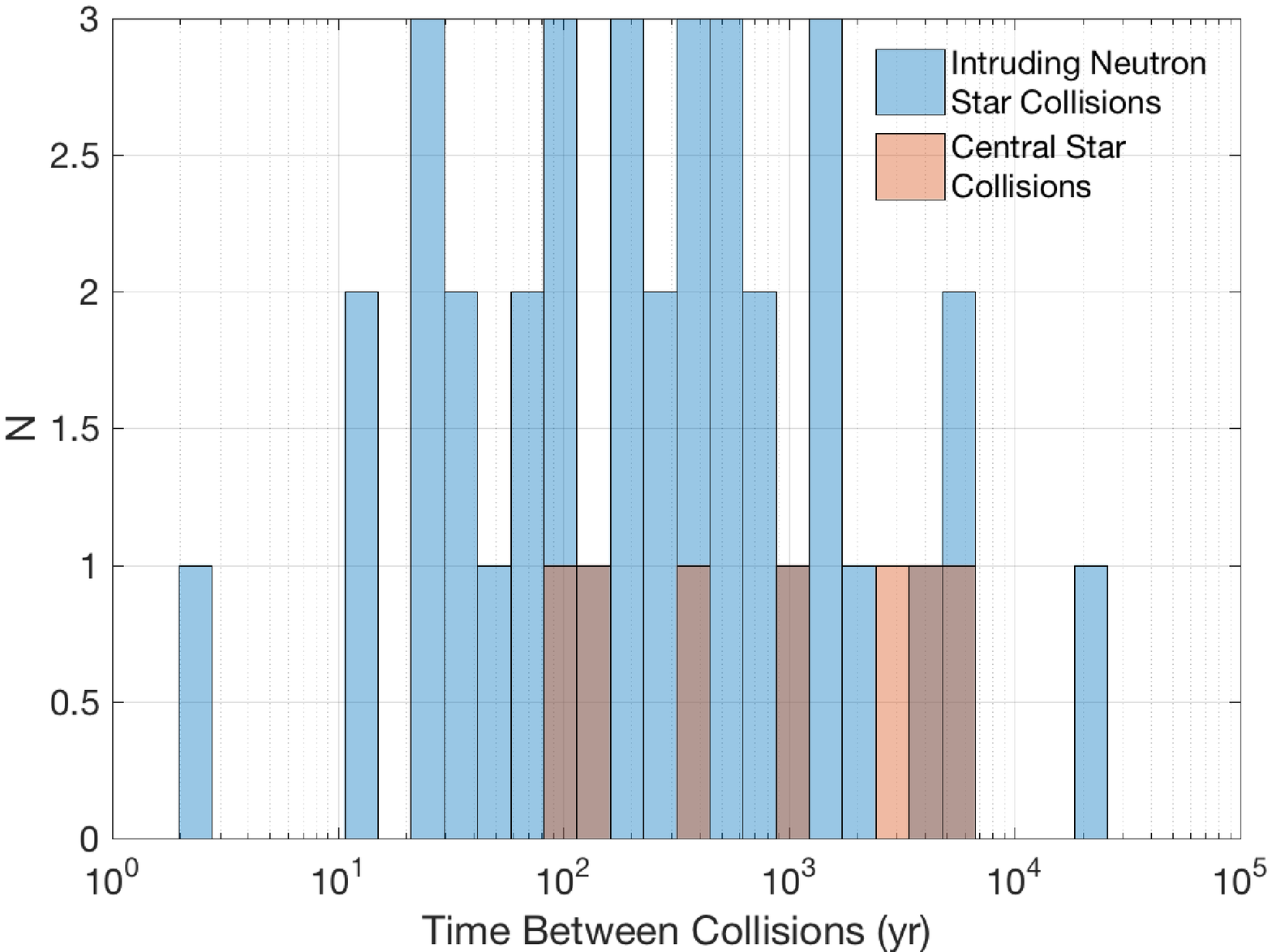}
\includegraphics[width=8.7cm]{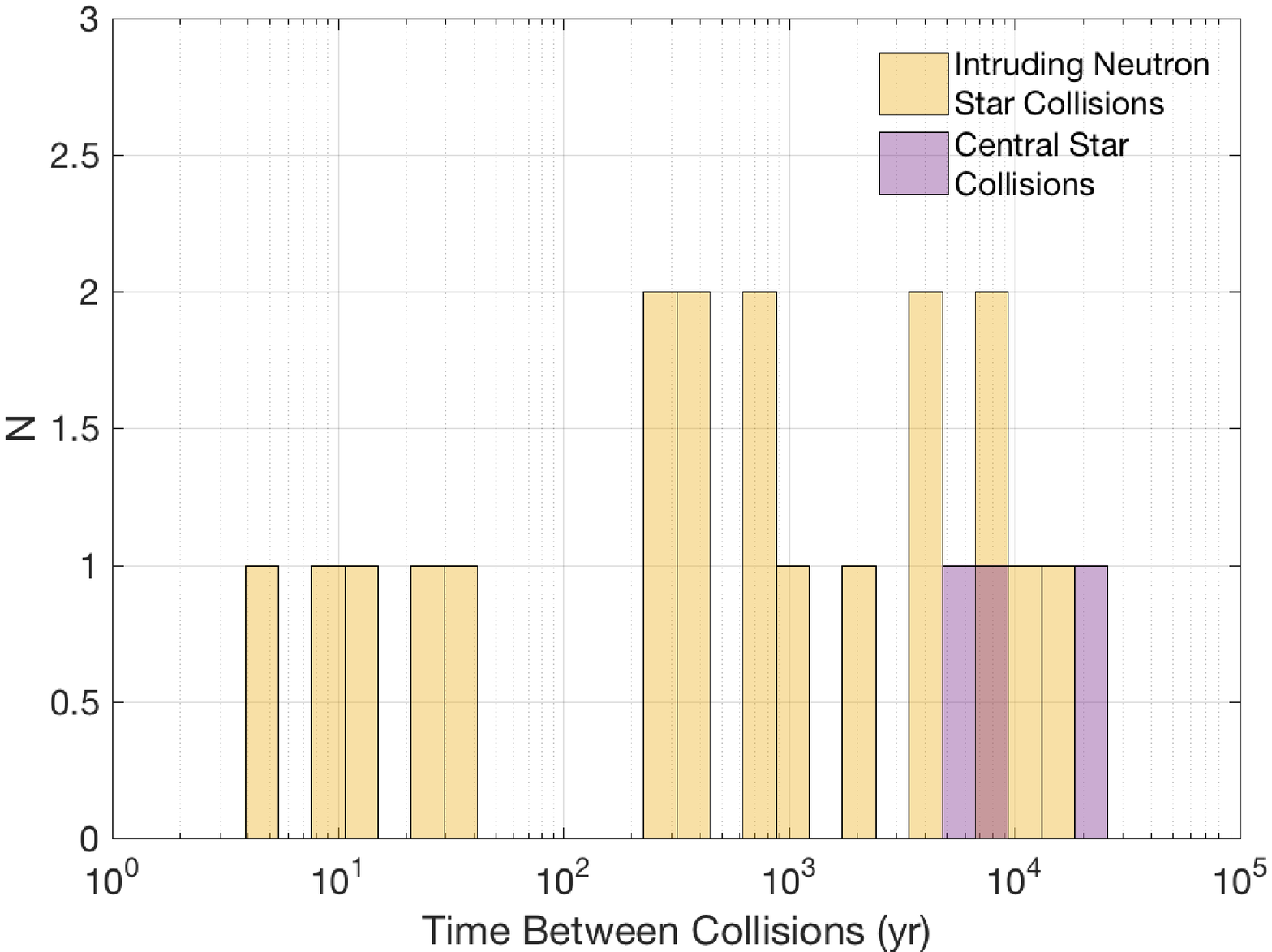}
\caption{The number of collisions as a function of time between each collision for the non-capture scenario (left panel) and for the capture scenario (right panel). In the left panel, the intruding NS is denoted with blue and central star with red. In the right panel, the intruding NS is denoted with yellow and central star with purple.}
\label{time_int}
\end{figure*}

The right panel of Fig.~\ref{setup} shows the final distribution of the surviving debris disc after a time of $100,000$ years. The majority of the debris belt becomes unstable except for a population that resides close to the central star. We show the eccentricity versus the semimajor axis distribution  of the test particle population at times $t= 0\, P_{\rm orb}$, $0.67\, P_{\rm orb}$, $1.67\, P_{\rm orb}$ and $t=166.67\, P_{\rm orb}$ shown in Fig.~\ref{e0p5}. The NS begins at apastron and has an orbital period of roughly $600\, \rm yr$. As the system evolves, the outer parts of the belt become unstable, increasing the eccentricity of the test particles. As the NS approaches periastron, the majority of the debris disc has already been scattered.  This unstable nature extends throughout the belt as time increases. The belt is stable close to the central star in $R \lesssim 15\, \rm au$. 

Next, we examine the scenario that resembles the NS being captured by a star with an debris belt. The left panel of Figure~\ref{e0p9} shows the initial setup for the NS capture model, while the right panel shows the final distribution of the debris belt. Much like the non-capture scenario, the belt becomes unstable as the NS approaches periastron. Figure~\ref{e0p9_dist} shows the eccentricity versus the semi--major axis distribution of the test particle population at times $t=0 \, P_{\rm orb}$ and $t=14.97 \, P_{\rm orb}$. Again, as the system evolves, the outer parts of the belt become unstable, increasing the eccentricity of the test particles. Next, we examine the impact rate of the test particles that have become unstable in each scenario.

\subsection{Numerical collision rate}

The fate of test particles with heightened eccentricities include impact with the central star or the NS, ejection from the system, or remains within the simulation domain. If a test particle collides with either of the stars, the test particle is considered impacted and removed from the simulation. Figure~\ref{rate} shows the impact rate onto the central star and onto the intruding NS in both non-capture (left panel) and capture (right panel) scenarios. We also show the time of first periastron approach for both models. Within both scenarios, the NS literally goes through the belt on the first periastron approach, however, there are only two collisions during the first periastron for the non-capture scenario. For the capture scenario, there is one collision during the first periastron passage. This is an interesting prediction, which states that the rate drops quickly with time, the highest being in the first orbit, but drops quickly in subsequent orbits. FRB 121102 has been observed for almost six years. It becomes active time and time again, which does not seem to be consistent with the prediction. However, since the orbital periods of the simulations are long, the source for FRB 121102 may still in the first encounter phase. In this case, we focus on the first encounter and comment on the deficiency of the rate (as above). In any case, the periodicity mentioned by \cite{Bagchi2017} should be irrelevant. Thus, a NS simply passing through a belt may not have a large amount of collisions. 

\cite{Dai2016}  used an asteroid belt analog as the source of debris. The numerical setup in this work made use of a larger Kuiper-belt analog. We now estimate the density of a Kuiper-belt analog that is able to produce the repetitive rate and then compare that with the densities of the current Kuiper belt and the primordial Kuiper belt. 

The observed rate of FRB 121102 during its active phase is about $1000\, \rm yr^{-1}$. According to our simulations, the total number of collisions onto the NS for each scenario is of the order of $10$ collisions per $100,000\, \rm yr$ with a disc number density of the order of $10^{-2}\, \rm au^{-3}$. To achieve the repetitive rate of $1000\, \rm yr^{-1}$, the density of our Kuiper belt analog would have to increase to $10^{7}\, \rm au^{-3}$. This density would predict $10^{10}$ collisions per $100,000\, \rm yr$, however, the velocity of the NS at periastron within our numerical simulations is two orders of magnitude smaller than the observed NS proper motion velocity, which is of order of $100\, \rm km/s$ (see section~\ref{velocity}). Since the rate is inversely proportional to $v_*$, this means that the numerical results overestimated the collision rate by two orders of magnitude. Thus, scaling our number density of the Kuiper-like belt by $9$ orders of magnitude would match the repetitive rate of $1000\, \rm yr^{-1}$. This density is three orders of magnitude greater than the current Kuiper belt and still an order of magnitude greater than the primordial Kuiper belt. Keep in mind that this scaled density is for a coplanar intruding NS to capture the highest rate of collisions. Realistically, the intruding NS would be misaligned to the plane of the debris belt and therefore the density of the belt would be greater than $10^{7}\, \rm au^{-3}$ to match the repetitive rate. Recall, that \cite{Dai2016} analytically found the number density to be $10^{9}\, \rm au^{-3}$ for an asteroid belt to match the repetitive rate. Thus, our numerical simulations suggest that a Kuiper belt analog could match the repetitive rate with a density greater than $10^7\, \rm au^{-3}$. If the debris disc was instead orbiting the intruding NS (i.e., the central star in our simulations), the rate of impacts would be much lower and the density required to match the observed repetitive rate would have to be larger than $10^8\, \rm au^{-3}$.

 We find another drawback to the collision model based on our numerical simulations. The repetitive rate of FRB 121102 is quite erratic, with a peak rate of about $3\, \rm hr^{-1}$ during its active phase \citep{Spitler2016,Scholz2016,Palaniswamy2016}. We explore our numerical results to identify if a short-time-scale erratic component is present. Figure~\ref{time_int} shows the number of collisions as a function of the time between each collision. The left panel shows the time interval distribution for the case where the NS eccentricity is $0.5$ and the right panel is when the NS eccentricity is $0.9$. For the former case, the distribution shows a close to one component Gaussian distribution with no short-time-scale erratic component. For the latter case, the distribution is also close to a one component Gaussian distribution. With more initial test particles, such a one-component Gaussian distribution may be enhanced without developing a short-time-scale erratic component.



\section{Conclusions}
\label{conc}
 We have examined  the FRB-asteroid collision model that has been postulated to explain the repeating FRB~121102. We summarize all the findings of the scenario below: 
\begin{itemize}
\item We first estimated the analytical rate of debris colliding onto a intruding NS with a density of a primordial Kuiper belt and with a low NS natal kick velocity. The primordial Kuiper belt is an extreme case since the current mass of the Kuiper belt is $1\%$ of its initial mass. Given this extreme case, the rate is still about three orders of magnitude lower than the observed rate of $3\, \rm h^{-1}$. This supports the findings of \cite{Dai2016}, that the source is most likely not located within a Milky Way analog, and that the potential progenitors could be in an extremely rare arrangement.

\item We find that the analytical duration to produce FRB by comets is consistent with the pulse width of FRB 121102 ($3 \pm 0.5 \, \rm ms$), assuming an average cometary nucleus radius of $5\, \rm km$. This suggests that a comet may be able to produce an FRB assuming that the long cometary tail does not disrupt the coherent emission needed to produce FRBs.

\item To compare our analytical interpretation to numerical integrations, we model a Kuiper-like debris disc around a central star with a NS on a highly eccentric orbits ($e=0.5$ and $e=0.9$). 
Within each scenario, the debris disc becomes unstable before the NS approaches periastron, which leads most comets to be scattered away from the belt rather than being accreted by the NS.


\item We estimate how dense our Kuiper-belt analog would have to be in order to reproduce the repetitive rate. We constrain the estimated density to be larger than $10^7\, \rm au^{-3}$ to match the observed repeating radio bursts for an intruding NS. If the disc happened to be around the NS, the density required would have to be larger than $10^8\, \rm au^{-3}$. These densities are $3-4$ orders of magnitude greater than the current Kuiper belt and $1-2$ orders of magnitude greater than the primordial Kuiper belt even if: (1) one introduces a Kuiper-belt like comet belt rather than an asteroid belt and assume that comet impacts can also make FRBs; (2) the NS moves $\sim 2$ orders of magnitude slower than their normal proper-motion velocity due to supernova kicks; and (3) the NS orbit is coplanar to the debris belt, which provides the highest rate of collisions.

\item Another drawback to this model is that the numerical simulations lack evidence for the erratic behavior of FRB 121102.

\end{itemize}
We conclude that if repeating FRBs are produced by comets colliding with an NS, the progenitor system must be in an extremely rare arrangement (i.e. an intruding NS plummeting through an extremely dense Kuiper-like comet belt or asteroid belt) to cause the repeating behavior as observed in FRB 121102. Thus, we do not rule out the mechanism proposed by \cite{Dai2016} but the evidence for such arrangements are sparse.   

\section*{Acknowledgements}
We thank Z. G. Dai for discussion and an anonymous referee for helpful suggestions. JLS acknowledges support from a graduate fellowship from the Nevada Space Grant Consortium (NVSGC). We acknowledge support from NASA through grants NNX17AB96G and NNX15AK85G. Computer support was provided by UNLV's National Supercomputing Center. 




\bibliographystyle{mnras}
\bibliography{main} 

\bsp	
\label{lastpage}
\end{document}